\documentclass{rmaa}

\title{The jet/counterjet symmetry of the HH~212 outflow}
\author{A. Noriega-Crespo
\affil{Space Telescope Science Institute}
A. C. Raga
\affil{Instituto de Ciencias Nucleares, UNAM}
V. Lora
\affil{Instituto de Radioastronom\'\i a y Astrof\'\i sica Te\'orica, UNAM}
J. C. Rodr\'\i guez-Ram\'\i rez
\affil{IAG, USP}
}

\fulladdresses{
\item A. Noriega-Crespo: Space Telescope Science Institute, 3700 San Martin Drive,
  Baltimore, MD 21218, USA
  (anoriega@stsci.edu)
\item A. C. Raga: Instituto de Ciencias Nucleares,
Universidad Nacional Aut\'onoma de M\'exico, Ap. 70-543, 04510 Cd. Mx., M\'exico
(raga@nucleares.unam.mx)
\item V. Lora: Instituto de Radioastronom\'\i a e Astrof\'\i sica Te\'orica,
  Universidad Nacional Aut\'onoma de M\'exico, Ap.3-72, 58089 Morelia, Michac\'an, M\'exico
\item J. C. Rodr\'\i guez Ram\'irez: Instituto de Astronomia, Geof\'\i sica e Ci\^encias
  Atmosf\'ericas,
  Universidade de S\~ao Paulo, R. do Mat\~ao 1226, 05508-090 S\~ao Paulo, SP, Brasil}

\shortauthor{Noriega-Crespo et al.}
\shorttitle{Jet/counterjet symmetry of HH~212}

\SetVolume{40} \SetFirstPage{001} \SetYear{2019}
\ReceivedDate{\today} 
\AcceptedDate{Year Month Day} 

\resumen{Presentamos im\'agenes del Spitzer (IRAC) y una imagen
  del VLT a 2.1~$\mu$m del flujo HH~212. Encontramos que
  este flujo tiene una fuerte simetr\'\i a, con pares de nudos en el jet/contrajet con
  differencias de posici\'on $\Delta x<1''$.
  Deducimos que los pares de nudos jet/contrajet son eyectados con diferencias de tiempo
  $\Delta \tau_0\sim 6$~yr y de velocidad $\Delta v_0\sim 2$~km~s$^{-1}$.
  Tambi\'en analizamos las desviaciones
  de las posiciones de los nudos perpendiculares al eje del flujo, y las interpretamos como resultado
  de un movimiento binario orbital de la fuente. A trav\'es de este modelo, deducimos una masa de
  $\sim 0.7$~M$_\odot$ para la fuente, y una separaci\'on de $\sim 80$~AU para la binaria (suponiendo
  masas iguales para sus dos componentes). Finalmente, usamos los datos de IRAC y la imagen
  del VLT a 2.1~$\mu$m para medir las velocidades de movimientos propios, obteniendo valores
  de 50 a 170~km~s$^{-1}$.}

\abstract{We present Spitzer (IRAC) images observations and a VLT 2.1~$\mu$m
  image of the HH~212 outflow. We find that
  this outflow has a strong symmetry, with jet/counterjet knot pairs with $\Delta x<1''$
  position offsets. We deduce that the jet/counterjet knots
  are ejected with time differences $\Delta \tau_0\sim 6$~yr and velocity differences
  $\Delta v_0\sim 2$~km~s$^{-1}$. We also analyze the deviations of the knot positions perpendicular
  to the outflow axis, and interpret them in terms of a binary orbital motion of the outflow source.
  Through this model, we deduce a $\sim 0.7$~M$_\odot$ mass for the outflow source, and a separation
  of $\sim 80$~AU between the components of the binary (assuming equal masses for the two components).
  Finally, using the IRAC data and the VLT 2.1~$\mu$m image we have measured the proper motion velocities,
  obtaining values from 50 to 170~km~s$^{-1}$.}

\keywords{shock waves --- stars: winds, outflows ---
Herbig-Haro objects --- ISM: jets and outflows --- ISM: kinematics and dynamics ---
ISM: individual objects (HH212) --- stars: formation}

\begin{document}
\maketitle

\section{Introduction}

The existence of symmetric emitting knots (at similar distances from the outflow source)
along some bipolar Herbig-Haro (HH) outflow systems imply highly synchronized jet/counterjet
ejections, and therefore a small spatial extent for the jet production region. This was
pointed out by Raga et al. (2011a) in their study of Spitzer Infrared Array Camera (IRAC) images of the
HH~34 outflow. In a second paper, Raga et al. (2011b) developed a ballistic jet model
which constrains the ejection asymmetries using the observed jet/counterjet
structures, and applied the model to the HH~34 and HH~111 outflows.

These analyses of jet/counterjet asymmetries have been carried out with IR Spitzer images
in the 4 IRAC channels (centered at 3.6, 4.5, 5.8 and 8.0 $\mu$m). This is because at
optical wavelengths, larger jet/counterjet asymmetries are found in the knots close
to the outflow source, many times with one of the two lobes being undetected
because of higher optical extinction (this is the case, e.g., of the HH~34, HH~111 and
HH~1/2 outflows). The intrinsic symmetry of the two lobes is then only visible
at infrared (IR) wavelengths, as first shown in a quite dramatic way by the H$_2$ 2.1~$\mu$m
observations of HH~111 of Gredel \& Reipurth (1994).

A clear candidate for this kind of study is the HH~212 outflow, which is
an impressive ``IR jet'', discovered at IR wavelengths by Zinnecker et al. (1998)
and with only very faint optical emission (Reipurth et al. 2019). This outflow
lies very close to the plane of the sky (Claussen et al. 1998) and is at
a distance of approximately 400~pc (Anthony-Twarog 1982, Kounkel et al. 2017).
Recent proper motion determinations (Reipurth et al. 2019) show that the
jet and the counterjet have a velocity $\approx 170$~km~s$^{-1}$.

H$_2$ 2.1~$\mu$m
observations of HH~212 (Davis et al. 2000, Smith et al. 2007, Correia et al. 2009) show
that this outflow has emitting structures with quite evident jet/counterjet symmetries.
We present Spitzer images in the four IRAC channels (I1-I4, at
3.6, 4.5, 5.8 and 8.0 $\mu$m) and an archival VLT 2.1~$\mu$m image of HH~212,
and determine the positions of knots along
the jet and the counterjet.

We then use the knots within $40''$ from the
outflow source to calculate jet/counterjet knot position offsets. These
offsets (as a function of distance from the outflow source) are then
interpreted in terms of the ballistic outflow model of Raga et al. (2011b)
in order to constrain the jet/counterjet asymmetries of the ejection
process. We also study the deviations of the knot positions perpendicular
to the outflow axis, and interpret them in terms of the ``orbiting outflow
source'' model of Masciadri \& Raga (2002).

The paper is organized as follows. The observations are discussed in section 2.
The measurement of knot intensities and positions (as well as the determinations
of jet/counterjet knot offsets) are presented in section 3. Section 4 presents interpretations
of the HH~212 measurements, including an application of the jet/counterjet asymmetry
model of Raga et al. (2011b), an application of the ``orbiting source jet'' model of
Masciadri \& Raga (2002),
an evaluation of the difference in extinction towards the jet and the counterjet, and
a discussion of the features of the outflow that do not show a clear jet/counterjet
symmetry. The results are summarized in section 5.

\begin{figure*}[!t]
\includegraphics[width=2\columnwidth]{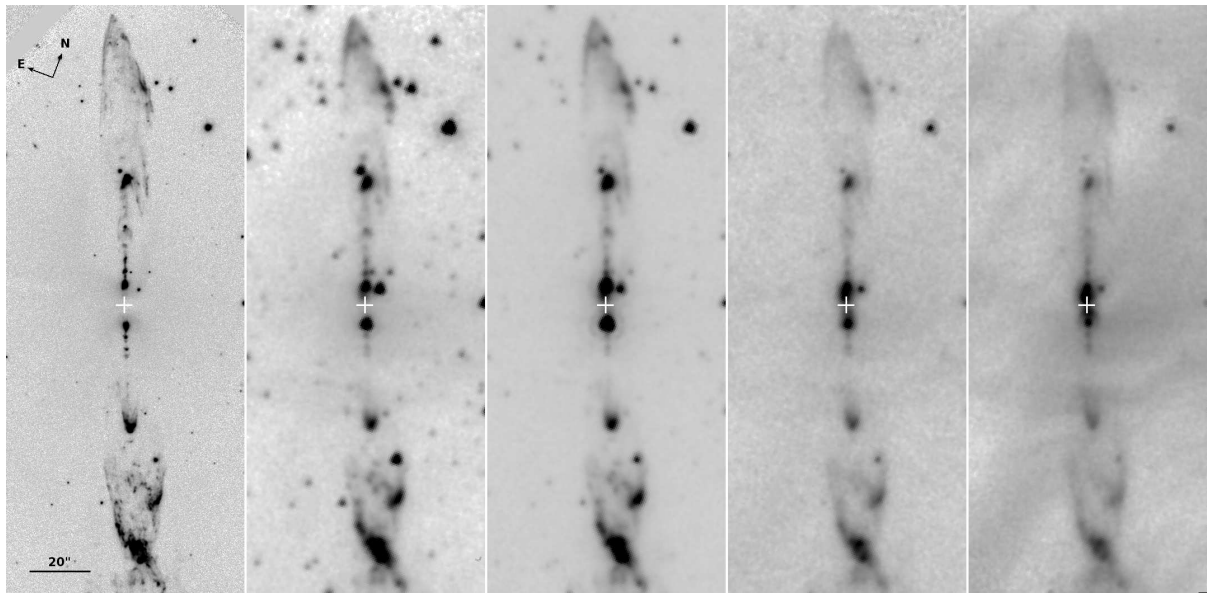}
\caption{The HH~212 outflow in H$_2$~2.1~$\mu$m (left frame) and in the four IRAC channels
  I1-I4, at 3.6, 4.5, 5.8 and 8.0~$\mu$m (four frames towards the right). The orientation and
  the scale of the images are shown on the left frame. The white cross (in all of the frames)
indicates the position of the outflow source.}
\label{fig1}
\end{figure*}

\section{The observations}

The  IRAC data was obtained during the Cryo-Spitzer mission, program PID 3315 (PI
Noriega-Crespo) on ``Emission from H2, PAHs and Warm
Dust in Protostellar Jets''. The data was collected in the four IRAC bands using a
30~sec high dynamic range (HDR) frame time and a 12 point medium scale
Reuleaux dither pattern. A small 2$\times$1 mosaic with a 260 arcsec stepsize  was
used to capture the outflow within the field of view (FoV) of the
four IRAC (3.6, 4.5, 5.8 and 8.0 $\mu$m) channels. The resulting images
have a $0.6''$ pixel size.

We have used the final reprocessing
from the Spitzer Archive with a standard angular resolution of
FWHM$\sim$2 arcsec. Figure 1 shows the entire outflow in the four bands.
As expected, the HH 212 jet itself is brighter
at 4.5 $\mu$m, given that some of the bright pure (0-0) rotational lines fall within
the IRAC Channel 2 bandpass, i.e. S(9) 4.6947, S(10) 4.4096
and S(11) 4.1810 $\mu$m (Noriega-Crespo et al 2004a, 2004b; Looney et al. 2007;
Tobin et al. 2007; Ybarra \& Lada 2009; Maret et al. 2009, Raga et al. 2011a;
Noriega-Crespo \& Raga 2012). {The IRAC Channel 2 map could also be brighter
because of the CO rovibrational lines that fall in its range. However these
lines require high temperatures and densities 
that do not normally occur in protostellar jets.}
Nevertheless, the jet is detected in all four bands
(Figure 1).

We also used a VLT H$_2$ 2.1~$\mu$m image obtained with the High Acuity Wide field K-Band Imager (HAWK-I), 
as part of its Science Verification program (PI Schneider, ``How symmetric is a symmetric flow. A deep H2 image 
of the Herbig Haro object 212'') observed in Janauary 2018, and enhanced by the ground-layer adaptive 
optics module (GRAAL) with an image quality of the order $0.2''$. The raw and reduced data are available through the
ESO archive. The uncalibrated image has a $0.106''$ pixel size (Leibundgut et al. 2018). This
image is shown in the left frames of Figures 1 and 2.

\begin{figure*}[!t]
\includegraphics[width=2\columnwidth]{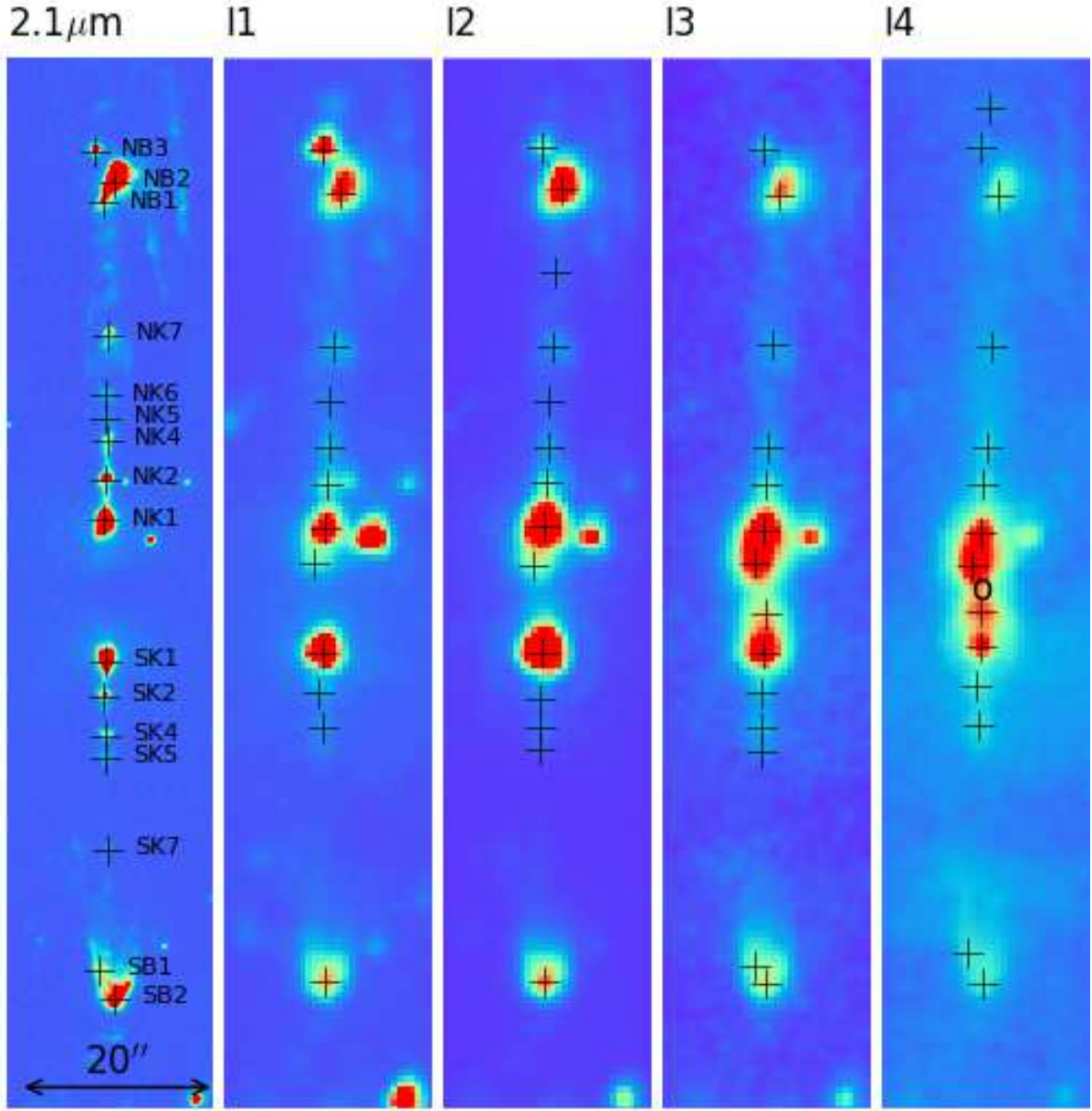}
\caption{The central region of the HH~212 outflow in in H$_2$~2.1~$\mu$m (left frame)
  and in the four IRAC channels (I1-I4,
  with 3.6, 4.5, 5.8 and 8.0~$\mu$m central wavelengths, respectively). The ordinate
  points to the NNE (at $22.5^\circ$ clockwise from N). The circle in the
  central region of the I4 map is the position of the source. The crosses indicate the
  positions of knots along the jet/counterjet (see the text). The knot identifications
  of Lee et al. (2007) are given on the H$_2$~2.1~$\mu$m map. The images are displayed
  with a linear colour scale.}
\label{fig2}
\end{figure*}

\section{The knot positions and intensities}

Figure 2 shows the H$_2$ 2.1$\mu$m and
the I1-I4 IRAC maps (with 3.6, 4.5, 5.8 and 8.0 $\mu$m
central wavelengths, respectively) of the HH~212 outflow. The images have
been rotated $22.5^\circ$ clockwise, so that the outflow axis is parallel to
the ordinate. The position of the outflow source (for which we have used
the position given in section 3.1 of Galv\'an-Madrid et al. 2004) is shown with
a black circle in the central region of the I4 map. On the H$_2$~2,1~$\mu$m map we show
the identifications given by Lee et al. (2007) for the H$_2$ knots.

In order to find the positions of the jet/counterjet knots, we have convolved
the I1-I4 maps with a ``Mexican hat'' wavelet with a central peak of
$\sigma=2$~pixel radius, which has the effect of isolating well defined
emission peaks from the spatially more extended emission. On these
convolved maps we search for peaks along the jet axis with an intensity
larger than a cutoff values $I_c$ (for which we have chosen values of
0.03, 0.05, 0.1 and 0.1 mJy/sterad for the I1, I2, I3 and I4 channels,
respectively), and carry out paraboloidal fits in
$3\times 3$ pixel regions (around each of the peaks) to determine
the knot positions. This procedure is described in detail by Raga
et al. (2017).

The H$_2$ 2.1$\mu$m jet/counterjet knot positions were found on a
convolution of this image with a central peak of $\sigma=5$~pixel radius.
We have selected peaks which have at least $10^{-2}$ times the peak knot
intensity (which is found for one of the SB knots of Lee et al. 2007)
in the convolved frame.

The resulting knot positions are shown as black crosses on the H$_2$ 2.1$\mu$m
and I1-I4 maps of Figure~2. It is clear that many of the knots along the NE jet
(top half of the frames) have corresponding emitting structures in the SW
counterjet (bottom half of the frames).

The knot located $\approx 26''$ to the N of the source
(labeled NK7 by Lee et al. 2007)
has no detectable counterpart in the counterjet in the IRAC I1-I4 maps.
However, in the H$_2$ 2.1$\mu$m image we do detect a faint counterpart along
the counterjet (labeled SK7).

In Figure 3 we show the peak intensities of the jet and counterjet
knots in the I1-I4 IRAC maps
(measured on the convolutions with a $\sigma=2$~pixel radius wavelet) as a function of
distance $x$ from the outflow source. This distance has been measured along the ordinate
of Figures 1 and 2, but (as the offsets of the knots along the abscissa are very small),
almost identical values are obtained if one takes the radial source/knot distances.

The jet and counterjet knots at similar
distances from the source have intensities that differ by factors of $\sim 2$.
We see that at similar distances from the source:
\begin{itemize}
\item in most cases the NE jet knots (in blue) are brighter than the SW counterjet
  knots (in red),
\item the ratios between the jet and counterjet knot intensities generally
  becomes smaller for the longer wavelength IRAC channels (see Figure 2).
\end{itemize}
These trends can be interpreted as the result of a different extinction
towards the two outflow lobes, as described in section 4.3.

In the top frame of Figure 4 we show the peak H$_2$ 2.1$\mu$m intensities of the jet and counterjet
knots (measured on the convolution with a $\sigma=5$~pixel radius wavelet) as a function of
distance $x$ from the outflow source. The intensities are given in units of the peak
intensity of the SK1 knot (see Figure 2). The jet and counterjet knots at similar
distances from the source have intensities that differ by factors of $\sim 3$,
except for knots NK7 and SK7 (at $\approx 26''$ from the source) which have
intensities that differ by a factor of $\sim 10$.

\begin{figure}[!t]
\includegraphics[width=\columnwidth]{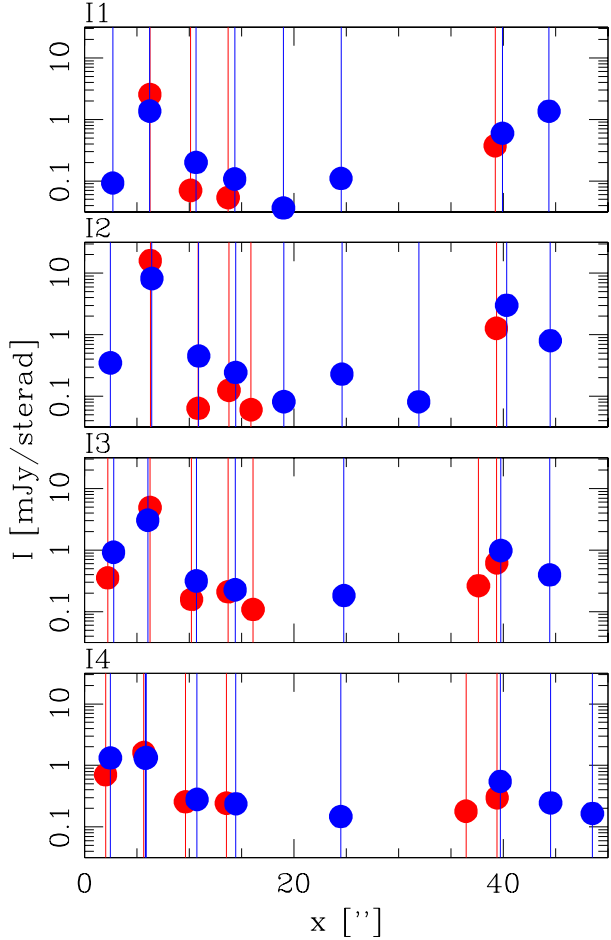}
\caption{Fluxes measured for the knots along the jet (blue circles with vertical lines)
  and counterjet (red circles) as a function of distance $x$ along the outflow axis
  in the I1-I4 IRAC channel maps. The fluxes
  are given in mJy/sterad. The vertical lines are shown so as to highlight the occurrence of
 jet/counterjet knot pairs with closely matched positions.}
\label{fig3}
\end{figure}

We now use the H$_2$ 2.1$\mu$m map (which has a higher angular resolution than
the IRAC maps) to define jet/counterjet knot associations with pairs of knots which
have values of $|x|$ (the distance to the outflow source) differing by
less than $2''$. For these pairs of knots, we calculate the
jet$-$counterjet knot position offsets $\Delta x=x_j-x_{cj}$ as a function of $x=x_j$.
In the bottom frame of Figure 4 we show the resulting $\Delta x$ (crosses) and $(\Delta x)^2$ (squares)
as a function of $x$. This plot shows a trend of marginally increasing jet-counterjet
knot position offsets with distance from the outflow source.

\begin{figure}[!t]
\includegraphics[width=\columnwidth]{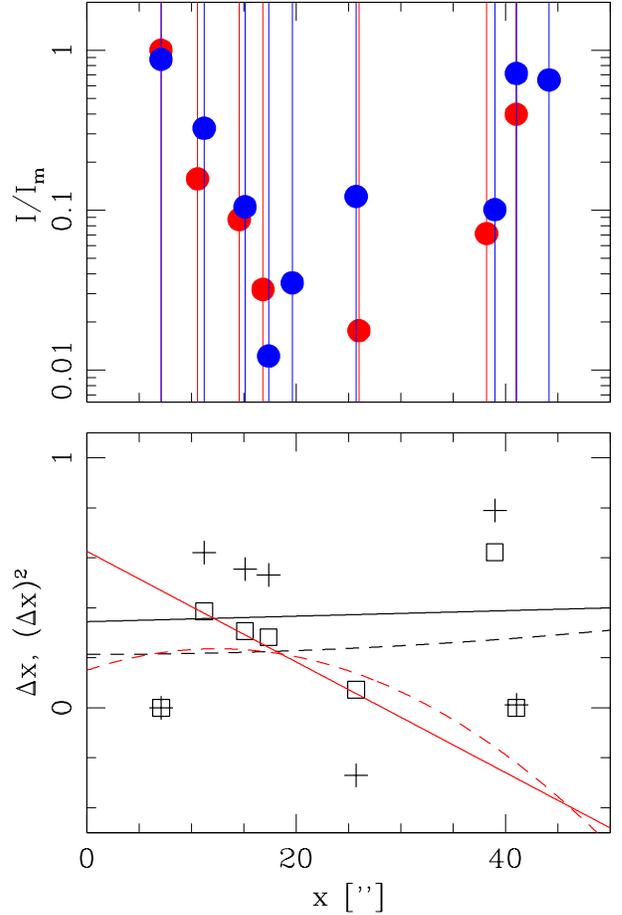}
\caption{Top frame: Fluxes measured for the knots along the jet (blue circles with vertical lines)
  and counterjet (red circles) as a function of distance $x$ along the outflow axis
  in the H$_2$ 2.1$\mu$m map. The fluxes are given in units of the flux of the SK1 knot.
  Bottom frame:
  jet/counterjet knot offsets $\Delta x$ (crosses) and the squares $(\Delta x)^2$
  of these values (squares) as a function of distance from the source. The results of the linear
  (black solid lines) and quadratic fits (black dashed lines) to the $\Delta x$ and $(\Delta x)^2$ vs. $x$
  dependencies are shown (see the text). The red lines are the corresponding fits to the points
  with $x<30''$ only. The jet/counterjet knot offsets have measurements
  errors of $\sim 0.05''$ (corresponding to $\sim 1/2$ pixel).}
\label{fig4}
\end{figure}

\begin{figure}[!t]
\includegraphics[width=\columnwidth]{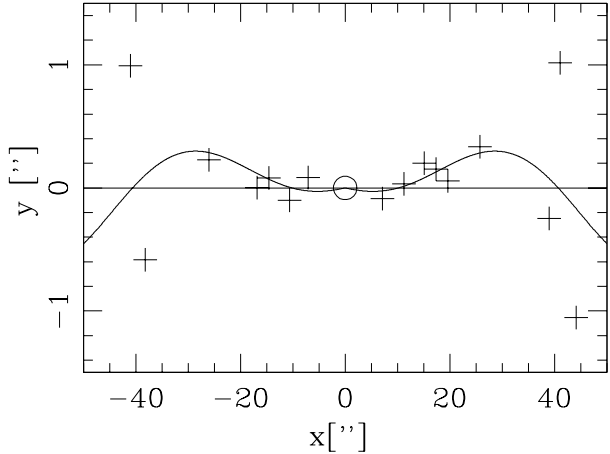}
\caption{Positions of the inner HH~212 knots measured on the H$_2$ 2.1$\mu$m frame.
  The $x$ coordinate lies along (positive $x$ to the N) and the
  $y$ coordinate across to the outflow axis (positive $y$ to the E). The solid
  curve is a least squares fit to the knots with $|x|<30''$ of the ballistic
  ``orbiting source jet'' of Masciadri \& Raga (2002), as described in the text. The
  positions along and across the outflow axis have errors of $\sim 0.03''$.}
\label{fig5}
\end{figure}

In Figure 5 we show the $(x,y)$ knot positions of the inner H$_2$ 2.1$\mu$m knots,
with $x$ measured along (positive values for the N jet) and $y$ across the outflow
axis (positive values to the E) from the position of the outflow source. It appears
that the knots with $|x|<30''$ have offsets (with respect to the outflow direction)
with a jet/counterjet mirror symmetric pattern. This result is discussed in more detail
below.

{Finally, we have used the new H$_2$~2.1$\mu$m image (obtained in January 2018)
together with the IRAC I1-I4 maps (obtained in February 2005) to estimate the proper motions
of the HH~212 knots within $50''$ from the outflow source. This of course gives only rough
estimates of the proper motions, as different knot morphologies in the different spectral
bands can in principle lead to position offsets that are not due to proper motions.

We have proceeded as follow. For the knots that are present in all of the I1-I4 IRAC maps,
we first calculate the average positions (along and across the outflow axis), and the
standard deviations of these positions. We then use these I1-I4 ``first epoch'' average knot
positions to calculate the knot proper motions together with their corresponding positions in the
H$_2$~2.1$\mu$m ``second epoch'' map. The proper motion velocities calculated from these
knot offsets (assuming a distance of 400~pc to HH~212) are given in Table~1.
\begin{table}[!t]
\small
\caption{Proper motions of the HH~212 knots}
\begin{center}
\begin{tabular}{rrrr}
\hline
\hline
Knot & $x$\phantom{0} & $v_x$\phantom{000} & $v_y$\phantom{000} \\
  & [$''$] & \multicolumn{2}{c}{[km  s$^{-1}$]} \\
\hline
NK1 &    7.1 &  141 (30) &   -3 (12) \\
NK2 &   11.2 &   69 (15) &   -1 (12) \\
NK4 &   15.1 &  105  \phantom{0}(7) &    1 (10) \\
NK7 &   25.7 &  170 (17) &  -28 \phantom{0}(7) \\
NB2 &   41.1 &  165 (36) &   10 (12) \\
NB3 &   44.2 &  -44 (10) &  -29 (17) \\
SK1 &   -7.1 & -145 (39) &   13 (21) \\
SK2 &  -10.6 &  -55 (64) &   48 (22) \\
SK4 &  -14.6 & -128 (13) &   14 (10) \\
SB2 &  -41.0 & -251 (11) &   32 (12) \\
\hline
\hline
\end{tabular}
\end{center}
\end{table}

This Table gives the knot identifications (shown in the left frame of Figure 2), the positions
$x$ along the outflow axis (measured in the H$_2$~2.1~$\mu$m map, with positive $x$ pointing
along the N jet), and the proper motion velocities along (positive values to the N) and across
(positive values to the W) the outflow axis with their errors (in parentheses). Even though
the errors shown are quite small for most of the
knots, it is likely that there are larger systematic errors due to the fact that
we compare images with different emission features.

The determined proper motion velocities are well aligned with the directions of the
jet and the counterjet axes, except for knot NB3. This knot has a motion directed to the NW,
which could indicate that it does not belong to the HH~212 outflow or that it has a substantially
different morphology in H$_2$~2.1~$\mu$m than in the other spectral bands.

Most of the knots have axial velocities in the range from $\sim 50$ to 170~km~s$^{-1}$, which is roughly
consistent with the previously determined proper motions of the HH~212 knots: $115\pm 50$~km~s$^{-1}$ by
Lee et al. (2015) and somewhat higher velocities by Reipurth et al. (2019). If one compares our results
(shown in Table~1) with Table~2 of Reipurth et al. (2019), one finds quite good agreements for the motions
of most of the knots present in both tables (note that the knot that we have labeled NB3 does
not correspond to the knot with the same denomination in Reipurth et al. 2019).}

\section{Interpretation of the results}

\subsection{The jet/counterjet knot position asymmetries}

We use the jet/counterjet knot offsets for constraining the jet/counterjet
asymmetries along the outflow axis shown in Figure 4. We do this using the model of Raga et al. (2011b).
In this model, one assumes that:
\begin{itemize}
\item the knots travel ballistically,
\item the jet/counterjet knot pairs are ejected with velocities that differ
  by $\Delta v$ (positive values indicating a faster knot along the jet), with this velocity difference
  following a uniform distribution with a mean value $v_0$ and a half-width $\Delta v_0$,
\item the knot pairs are ejected with a time-difference $\Delta \tau$ (positive values
  indicating an earlier jet knot ejection), with the time-difference following a uniform
  distribution with mean value $\tau_0$ and a half-width $\Delta \tau_0$.
\end{itemize}
The values of $v_0$, $\Delta v_0$, $\tau_0$ and $\Delta \tau_0$ can then be determined
by carrying out a linear fit to the $\Delta x$ vs. $x$ trend and a quadratic fit
to the $(\Delta x)^2$ vs. $x$ trend observed in a particular jet/counterjet system. The
values of the mean values and half-widths of the ejection velocity and time distributions
can be found from the coefficients of these fits using equations (4) and (6)
of Raga et al. (2011b).

We carry out the linear and quadratic fits to the $\Delta x$ vs. $x$ and the $(\Delta x)^2$ vs.
$x$ values (respectively) obtained from the H$_2$ 2.1$\mu$m map. The results of these fits
are shown with solid and dashed lines, respectively, in the bottom frame
of Figure 4. With the fitting
coefficients we determine the characteristics of the asymmetrical jet/counterjet ejection
time and velocitiy distributions (see above and Raga et al. 2011b):
\begin{itemize}
\item $v_0=(0.16\pm 0.17)$~km~s$^{-1}$, $\Delta v_0=(1.60\pm 0.12)$~km~s$^{-1}$,
\item $\tau_0=(4.38\pm 1.54)$~yr, $\Delta \tau_0=(6.77\pm 4.74)$~yr.
\end{itemize}
These parameters were derived assuming a distance of 400~pc and a flow velocity
of $(170\pm 30)$~km~s$^{-1}$ for HH~212 (see Reipurth et al. 2019).

In other words, the jet/counterjet knot position asymmetries of HH~212
can be explained with:
\begin{itemize}
\item an ejection velocity asymmetry with a
  distribution centered at 0 (i.e., the value of $v_0$ determined
  from the fits is not significantly different from 0, see above) and half-width
  of $\approx 1.6$~km~s$^{-1}$,
\item an ejection time asymmetry with a distribution centered at
  $\approx 4$~yr, and a (barely significant) width of $\approx 7$~yr.
\end{itemize}
These results are qualitatively similar
to the ones found for the HH~34 jet/counterjet system by Raga et al. (2011b).

It is fair to say that through this analysis we are basically not detecting a
significant asymmetry in the jet/counterjet ejections, and are only estimating
upper boundaries (of $\sim 2$~km~s$^{-1}$ for the velocity and $\sim 4$~yr for
the ejection time) for possible asymmetries in the ejections.

{The large uncertainty in our estimate of the ejection asymmetries is illustrated
with the following exercise. One could argue that
the local intensity maxima of knots NB1, NB2, SB1 and SB2 actually correspond
to local features in larger bow shocks (see Figure 2), and therefore the associations
NB1-SB1 and NB2-SB2 used to calculate the offsets at $x\approx 40''$ (see the two
frames of Figure 4) are not necessarily meaningful. Therefore, we repeat the
linear and quadratic fits (to the $\Delta x$ and $(\Delta x)^2$ vs. $x$
dependencies) only using the knot offsets obtained for $x<30''$.

The results of these fits are shown with the solid (linear fit
to $\Delta x$ vs. $x$) and dashed (quadratic fit to $(\Delta x)^2$ vs. $x$)
red lines in the bottom frame of Figure 4. These fits
do not yield physical estimates of the ejection variability, as the formalism
of Raga et al. (2011b) gives complex values for the derived parameters for
the ejection distributions when using the resulting values of the coefficients
of these fits.

Given the lack of a clear correlation of the jet/counterjet knot offsets as
a function of distance from the source (evidenced by the fact that the results
change in a quite drastic way by removing the knots at $x\sim 40''$) it is
probably fairer to just note that the knot offsets have a mean value
$\overline{|\Delta x|}=(0.40\pm 0.29)''$. This corresponds to an average
time-difference $\overline{\Delta \tau}=\overline{|\Delta x|}/v_j=(4.5\pm 3.3)$~yr
(for a distance of 400~pc and $v_j=170$~km~s$^{-1}$, see above).
This estimate is consistent with the $\Delta \tau_0=(6.77\pm 4.74)$~yr width
for the ejection time distribution deduced above using the formalism of
Raga et al. (2011b) and the fit to all of the knot offsets shown in the bottom
frame of Figure 4.}

\subsection{The mirror symmetric precession pattern}

The inner jet/counterjet knot positions of HH~212 show sideways deviations
from the outflow axis with an apparent ``mirror symmetric pattern''. In Figure
5, we see that the jet/counterjet knots within $30''$ from the source show
trends of larger values of $y$ (i.e., towards the E) with increasing distances
from the source. At $x\approx \pm 40''$ we see the NB and SB knots (respectively),
which show a larger spread of $y$ values, as a result of the larger size of
the NB and SB structures.

The simplest explanation of mirror symmetric patterns in jet/counterjet systems
is in terms of an orbital motion of the (binary) outflow source. A ballistic, analytic model
of this situation was presented by Masciadri \& Raga (2002, for the case of a circular
orbit) and by Gonz\'alez \& Raga (2004, for elliptical orbits).

Noriega-Crespo et al. (2011) used the ``circular orbit model'' of Masciadri \&
Raga (2002) to fit the ``mirror symmetric precession pattern'' of the
HH~111 jet/counterjet system. From this fit, they derived estimates of
the orbital parameters and stellar masses of the assumed binary source of
the HH~111 system.

We follow these authors, and carry out a least squares fit of
the jet/counterjet locci:
\begin{equation}
  y=\kappa x \sin\left(\frac{2\pi}{\tau_o v_j}x-\psi\right)\,,
  \label{yx}
\end{equation}
\begin{equation}
  z=\kappa x \cos\left(\frac{2\pi}{\tau_o v_j}x-\psi\right)\,,
  \label{zx}
\end{equation}
where $x$ is the axial coordinate, and $(y,z)$ are the axes parallel to the orbital plane
(with $y$ being parallel to the plane of the sky). Also, $\psi$ is the orbital phase,
$\tau_o$ the orbital period, and $\kappa=v_o/v_j$ is the ratio between the orbital
and jet velocities. These equations (see Noriega-Crespo et al. 2011) correspond to the
small orbital radius limit of the circular orbit model of Masciadri \& Raga (2002).
It has also been assumed that the outflow is ejected perpendicular to the orbital plane.

We project equations (\ref{yx}-\ref{zx}) onto the plane of the sky assuming an angle
of $5^\circ$ between the outflow axis and this plane (see Reipurth et al. 2019),
and carry out a least squares fit to the mirror symmetric pattern of the knots within $30''$
of the HH~212 source. From this fit, we obtain:
\begin{itemize}
\item $\kappa=0.011\pm 0.001$, corresponding to an orbital velocity $v_o=(1.87\pm 0.17)$~km~s$^{-1}$
  for the $v_j=170$~km~s$^{-1}$ proper motion velocity of Reipurth et al. (2019),
\item $\tau_o=(638\pm 241)$~yr, where we have also assumed a distance of 400~pc to HH~212.
\end{itemize}
With these derived values for the orbital velocity and the orbital period, we can
derive the orbital radius:
\begin{equation}
  r_1=\frac{v_o\tau_o}{2\pi}=(40\pm 15)\,{\rm AU}\,,
  \label{r1}
\end{equation}
and a mass
\begin{equation}
  \frac{\alpha^3}{(1+\alpha)^2}M_1=\frac{\tau_o v_o^3}{2\pi G}=(0.168\pm 0.068)\,{\rm M_\odot}\,,
  \label{m1}
\end{equation}
for the primary (jet source) star. In equation (\ref{m1}), $G$ is the gravitational
constant $\alpha=M_2/M_1$ is the mass ratio of the binary. If we have an equal
mass binary (with $\alpha=1$), we then obtain a mass $M_1=(0.67\pm 0.27)$~M$_\odot$
and a binary separation of $2r_1=(80\pm 30)$~AU.

{We should note that Lee et al. (2015) made a fit of a much tighter precession
  spiral (with a spatial wavelength $\sim 6''$, corresponding to a $\sim 90$~yr period)
  to the observed CO/SiO emission of HH~212. This small scale structure of curved jet
  segments can also be seen in the NK1-NK4 region of the H$_2$~2.1~$\mu$m jet shown
  in the left panel of Figure 2. Analogously, the larger scale structure (with a
  period of $\sim 600$~yr, see above) we are describing here is also seen in the CO/SiO
  map shown in Figure 8 of Lee et al. (2015).}

\begin{figure}[!t]
\includegraphics[width=\columnwidth]{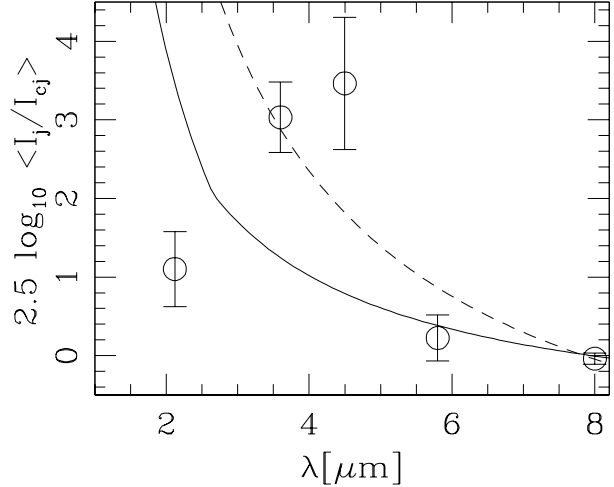}
\caption{Mean values of the jet/counterjet intensity ratio (calculated with
  the three knot pairs at $x=5\to 15''$, see the text) as a function of
  central wavelength of the IRAC channels and of the VLT H$_2$ map.
  The intensity ratios are shown as
  a magnitude, and as a function of the central wavelength. The
  solid line shows a least squares fit to the observed line ratios of a
  linearly transformed $R=5$ extinction curve. The fit to the data gives
  a higher absorption towards the counterjet of $A_v=16.9\pm 2.3$. The dashed
  line shows a fit to the intensity ratios with $\lambda>3\mu$m}
\label{fig6}
\end{figure}

\subsection{The jet/counterjet extinction}

As discussed in section 3, the jet knots (blue points) in the IRAC channel maps
are generally brighter than the counterjet knots (red points, see Figure 3) at the
same distances from the outflow source, especially for the shorter
wavelengths channels (I1 and I2). In order to quantify this effect,
we calculate the $I_j/I_{cj}$ jet-to-counterjet knot intensity ratios
for the three knots at distances $x=5\to 15''$ (from the outflow
source), and use these ratios to calculate a mean ratio
$<I_j/I_{cj}>$ for each of the four IRAC channels. Figure 6 shows the resulting
mean jet-to-counterjet $<I_j/I_{cj}>$ ratios (shown as magnitudes) as a function
of $\lambda$ (where $\lambda$ is the central wavelength of the
four IRAC channels). In this Figure we also show the jet-to-counterjet intensity
ratio of the knots seen in the 2.1$\mu$m map.

We have carried out a weighted least squares fits with a linear transformation
of the $R=A_v/E(B-V)=5$ extinction curve
(appropriate for star formation regions) of Fitzpatrick (1999), and we show the
results in Figure 6. We show two fits:
\begin{itemize}
\item a fit to the jet-to-counterjet intensity ratios measured in the IRAC
  maps (dashed curve in Figure 6). The fit gives a visual extinction to the counterjet higher by
  $A_v=16.9\pm 2.3$ than the extinction towards the jet,
\item a fit to the jet-to-counterjet intensity ratios measured in the IRAC
  maps and in the 2.1~$\mu$m map (solid curve in Figure 6). The fit gives a visual
  extinction to the counterjet higher by
  $A_v=7.2\pm 1.3$ than the extinction towards the jet.
\end{itemize}
These of course are estimates
only of the difference between the jet and counterjet extinction, and not
a determination of the value of the total extinction to the HH~212 outflow.

The measured $I_j/I_{cj}$ jet-to-counterjet knot intensity ratios shown in Figure 6
have quite large deviations from the extinction curves. This indicates that the
jet and counterjet knots at similar distances from the source have relatively large
intrinsic intensity differences, not attributable in a direct way to a difference
in the extinction.

\subsection{Knot NK7}

Knot NK7 is located along the NE jet at a distance $x\approx 25''$ from the outflow
source (see Figures 2 and 3). Lee et al. (2007)
show that this knot has a very faint H$_2$ 2.1~$\mu$m southern counterpart,
but they do not detect it in SiO and CO (at sub-mm wavelengths). We also
see the faint SK7 counterpart to NK7 in our H$_2$ image (see Figure2).
This result, together with the fact that we do not see the southern counterpart of NK7
in the IRAC images, indicates that this knot is intrinsically much brighter
along the NE jet than the coresponding ejection along the counterjet, and that this
strong brightness asymmetry is not an extinction effect (as the extinction should be
much less important at longer wavelengths).

Should we therefore conclude that even though the jet/counterjet ejection
from the HH~212 source appears to have a remarkable degree of symmetry (see
section 4.1), every now and then it produces highly asymmetrical ejections?
This is a possible interpretation of the lack of a bright counterpart
for the NK7 knot.

Another possible mechanism for producing the observed intensity asymmetry is that
knot NK7 corresponds to the merger of two knots (travelling down
the jet at slightly different velocities), and that the brightening
is associated with the merging process (which produces a knot
of boosted shock velocities). If this were the case, we might expect
to see a sudden brightening of an ``SK7'' knot (at $x\approx 25''$ from
the source) along the counterjet when the corresponding knot merger
occurs in the counterjet.

\section{Conclusions}

We present Spitzer (IRAC) observations and an H$_2$~2.1$\mu$m image
of the HH~212 outflow. In these maps,
the general structure of the two outflow lobes is seen (see Figure 1).

For the inner $\sim 1'$ of the outflow, we have determined the positions of
knots along the NE jet and SW counterjet (see Figure 2), and find that they
mostly fall into ``jet/counterjet knot pairs'' (with distances from the source
differing by at most $\sim 1.2''$). We then calculate the jet/counterjet
knot offsets $\Delta x$ as a function of distance $x$ from the outflow source
(see Figure 4). We carry out the analysis of knot position offsets with the
2.1~$\mu$m map, which has higher angular resolution than the IRAC maps.

We interpret the observed jet/counterjet position offsets with the
quasi-symmetric ballistic ejection model of Raga et al. (2011b). Through
this exercise we determine that the knot pairs are ejected with time-differences
$\Delta \tau_0\sim 6$~yr and velocity differences $\Delta v_0\sim 2$~km~s$^{-1}$.
These results are similar to the ones obtained for HH~34 by Raga et al. (2011b).
Clearly, an appropriate ejection model should have this degree of jet/counterjet
coordination.

One can in principle use the determined jet/counterjet ejection coordination
to estimate a physical size for the jet production region. In the cool, magnetized
ejection mechanisms appropriate for low mass young stars, the signal transmission
velocity (which could be either the Alfv\'en or the sound speed) is expected
to lie in the $v_s\sim 0.1\to 10$~km~s$^{-1}$ range. We would then predict a size
of $L=\Delta \tau_0 v_s \sim 0.1\to 10$~AU for the jet production region. The
lower limit of this size range is in agreement with the estimation of Lee et al. (2017)
of a $\sim 0.1$~AU size for the HH~212 outflow collimation region.

{We have used the knot positions measured on the H$_2$~2.1~$\mu$m image together with
  the IRAC maps (which were obtained $\approx 13$~yr earlier) to determine proper motions
  of the knots along the jet and the counterjet. We find generally good agreement with
  the proper motions obtained by Reipurth et al. (2019) with two H$_2$~2.1~$\mu$m
  epochs covering an $\approx 8$~yr time-interval.}

We have also analyzed the deviations of the knot positions perpendicular to the
mean axis of the outflow. These deviations show a mirror symmetric jet/counterjet
pattern, which can be interpreted in terms of a ballistic outflow from a source
in an orbit around a binary companion. We have fitted the model of an outflow
source in a circular orbit of Masciadri \& Raga (2002) to the observed deviations
(see Figure 5). From the model fit we deduce an $(80\pm 30)$~AU binary separation and
a $(0.67\pm 0.27)$~M$_\odot$ mass for the outflow source (assuming that the binary
companion has the same mass as the outflow source). This estimate for the separation
between the binary components coincides with the $\sim 90$~AU radius of the disk around the HH~212
source observed by Codella et al. (2014). Our estimated mass is somewhat
larger than the $\sim 0.15$~M$_\odot$ mass estimated by Lee et al. (2006, from observations
of an infalling envelope) and the $\sim 0.3$~M$_\odot$ mass estimated by Codella
et al. (2014, from the rotation of the disk) for the HH~212 outflow source.

The general structure of HH~212 has an important asymmetry in that the NK7 knot
(at $\sim 25''$ along the NE jet, see Figures 2 and 3) does not have a comparably
bright counterpart along the counterjet. This asymmetry is observed at all wavelengths
at which the HH~212 outflow has been observed (see Lee et al. 2007), and therefore
cannot be accounted for by a differential extinction (see sections 4.3 and 4.4). We suggest
that the asymmetric knot NK7 could be interpreted as a true ejection asymmetry, or
as a recent brightening of the knot due to the merger of two ``outflow events''. If
this latter explanation is correct, we might expect a future brightening of a
counterjet knot at a comparable distance to the outflow source.

Finally, we have used the wavelength dependence of the jet/counterjet intensity
ratio (measured in the four IRAC channels) to determine the difference in the
extinction to the HH~212 jet and counterjet. We conclude that the extinction towards
the counterjet is higher (than the one towards the jet) by $A_v\approx 10$~magnitudes.
This result is similar to the one found by Raga et al. (2019) for the HH~34 outflow.
However, we find large deviations between the extinction curve and the jet/counterjet
intensity ratios (as a function of wavelength). This indicates that
the jet and counterjet knots at similar distance from the outflow source have
relatively large intrinsic intensity differences.

\acknowledgments
AR acknowledges support from the DGAPA-UNAM grant IG100218. JCRR acknowledges
the Brazilian agency FAPESP grant 2017/12188-5. We thank an anonymous referee
for helpful comments, which gave rise to section 4.3.


\begin{thebibliography}

\bibitem{} Anthony-Twarog, B. J. 1982, AJ, 87, 1213

\bibitem{} Chini, R., Reipurth, B., Sievers, A., Ward-Thompson, D., Haslam,
  C. G. T., Kreysa, E., \& Lemke, R. 1997, A\&A, 325, 542

\bibitem{} Codella, C., Cabrit, S., Gueth, F. et al. 2014, A\&A, 568, L5

\bibitem{} Correia, S., Zinnecker, H., Ridgway, S. T., \& McCaughrean, M. J. 2009,
  A\&A, 505, 673

\bibitem{} Davis, C. J. Berndsen, A., Smith, M. D., Chrysostomou, A., \& Hobson, J.
  2000, MNRAS, 314, 241

\bibitem{} Fitzpatrick, E. L. 1999, PASP, 111, 63

\bibitem{} Galv\'an-Madrid, R., \'Avila, R., \& Rodr\'\i guez, L. F.
  2004, RMxAA, 40, 31

\bibitem{} Gonz\'alez, R. F., \& Raga, A. C. 2004, RMxAA, 40, 61
  
\bibitem{} Gredel, R., \& Reipurth, B. 1994, A\&A, 289, L19

\bibitem{} Kounkel, M., Hartmann, L., Loinard, L. et al. 2017, ApJ, 834, A142

\bibitem{} Lee, C.-F., Ho, P. T. P., Beuther, H. et al. 2006, ApJ, 639, 292
  
\bibitem{} Lee, C.-F., Ho, P. T. P., Hirano, N. et al. 2007, ApJ, 659, 499

\bibitem{} Lee, C.-F., Hirano, N., Zhang, Q. et al. 2015, ApJ, 805, 186

\bibitem{} Lee, C.-F.; Ho, P. T. P.; Li, Z.-Y. et al. 2017, Nature Astr., 1, 0152

\bibitem{} Leibundgut B. et al. 2018, Messenger, 172, 8
  
\bibitem{} Looney, L. W., Tobin, J. J. \& Kwon, W. 2007, ApJ, 670, 131

\bibitem{} Masciadri, E., \& Raga, A. C. 2002, ApJ, 568, 733

\bibitem{} Maret, S. et al. 2009, ApJ, 698, 1244

\bibitem{} Noriega-Crespo, A. et al. 2004a, ApJS, 154, 352

\bibitem{} Noriega-Crespo, A. et al. 2004b, ApJS, 154, 402

\bibitem{} Noriega-Crespo, A., Raga, V., Stapelfeldt, K. R., \& Carey, S. J.
  2011, ApJ, 732, L16

\bibitem{} Noriega-Crespo, A., \& Raga, A. C. 2012, ApJ, 750, 101 

\bibitem{} Raga, A. C., Noriega-Crespo, A., Lora, V., Stapelfeldt, K. R., \& Carey, S. J.
  2011a, ApJ, 730, L17

\bibitem{} Raga, A. C., Noriega-Crespo, A., Rodr\'\i guez-Ram\'\i rez, J, C., Lora, V.,
  Stappelfeldt, K. R., \& Carey, S. J. 2011b, RMxAA, 47, 277

\bibitem{} Raga, A. C., Noriega-Crespo, A., Rodr\'\i guez-Gonz\'alez, A., Lora, V.,
Stapelfeldt, K. R. \& Carey, S. J. 2012, ApJ, 748, 103

\bibitem{} Raga, A. C., Reipurth, B., \& Esquivel, A., et al. 2017, RMxAA, 53, 485

\bibitem{} Raga, A. C., Reipurth, B., \& Noriega-Crespo, A. 2019, RMxAA, in press

\bibitem{} Reipurth, B., Davis, C. J., Bally, J., Raga, A. C., Bowler, B. P.,
  Geballe, T. R., Aspin, C., \& Chiang, Hsin-Fang 2019, AJ, submitted

\bibitem{} Smith, M. D., O'Connell, B., \& Davis, C. J. 2007, A\&A, 466, 565
  
\bibitem{} Tobin, J. J., Looney, L. W., Mundy, L. G., Kwon, W. 
\& Hamidouche, M. 2007, ApJ, 659, 1404 

\bibitem{} Ybarra, J. E., \& Lada, E. A. 2009, ApJ, 695, 120 

\bibitem{} Zinnecker, H., Bastien, P., Arcoragi, J.-P., \& Yorke, H. W.
  1992, A\&A, 265, 726


  
\end{thebibliography}
\end{document}